\documentclass[12pt]{article}
\pdfoutput=1
\usepackage{amsmath}
\usepackage{amsthm}
\usepackage{graphicx,psfrag,epsf}
\usepackage{enumerate}
\usepackage{natbib}
\usepackage{color}

\newcommand{\blind}{0}

\newcommand{\real}{I\!\!R}  
\newcommand{\estim}{\widehat}
\def\Var{{\bf{Var}}}

\bibpunct{(}{)}{;}{a}{}{,}

\date{February 7, 2015}

\begin{document}

\def\spacingset#1{\renewcommand{\baselinestretch}%
{#1}\small\normalsize} \spacingset{1}


\if0\blind
{
  \title{\bf Estimation Stability with Cross Validation (ESCV)}
  \author{Chinghway Lim 
\\
    {Department of Statistics and Applied Probability}\\
    National University of Singapore
    \and
    Bin Yu \\
    Departments of Statistics and EECS\\
     University of California, Berkeley}
  \maketitle
} \fi

\if1\blind
{
  \bigskip
  \bigskip
  \bigskip
  \begin{center}
    {\LARGE\bf Title}
\end{center}
  \medskip
} \fi

\bigskip
\begin{abstract}
{Cross-validation ($CV$) is often used to select the regularization parameter in high dimensional problems. However, when applied to the sparse modeling method Lasso, 
 $CV$ leads to models that are unstable in high-dimensions,
and consequently not suited for reliable interpretation.
In this paper, we propose a model-free criterion $ESCV$ based on a new \emph{estimation stability (ES)} metric and $CV$. 
Our proposed $ESCV$ finds a smaller and locally $ES$-optimal model 
smaller than the $CV$ choice so that the it fits the data
and also enjoys estimation stability property.
We demonstrate that $ESCV$ is an effective alternative to $CV$ at a similar easily parallelizable computational cost.
In particular, we compare the two approaches with respect to several performance measures when applied to the Lasso on both simulated and real data sets. 
For dependent predictors common in practice, our main finding is that,
$ESCV$ cuts down false positive rates often by a large margin, while
sacrificing little of true positive rates. $ESCV$ usually outperforms
$CV$ in terms of parameter estimation while giving similar performance as $CV$ in terms of prediction.
For the two real data sets from neuroscience and cell biology, 
the models found by $ESCV$ are less than half of the model sizes by $CV$, but preserves $CV$'s predictive performance and corroborates with subject knowledge and independent work.
We also discuss some regularization parameter alignment issues that come up in both approaches. Supplementary materials are available online.}
\end{abstract}

\noindent%

\spacingset{1.45}

\section{Introduction}

\subsection{Regularization Methods}
\label{SecReg}
There is an ever increasing amount  of data in all fields of science and engineering.
Often, this data comes in high dimensions relative to the sample size, posing a new challenge to scientists, engineers, and decision makers.
These problems, plagued by the curse of dimensionality, suffer from overfitting when classical methods are applied.
Regularization methods are used to tackle this problem of overfitting head on, usually by imposing a penalty on the complexity of the solution or through early stopping. For example, in fitting the usual linear regression model, the Lasso \citep{Tib96} and ridge regression \citep{Tik43, Hoe62} adds a $L_1$ and $L_2$ penalty on the coefficient estimates respectively to the usual least squares fit objective function. Regularization methods can also take the form of early stopping iterative algorithms like classical forward selection or $L_2$-Boosting \citep{Fri01, BuhYu03, ZhaYu05, Zha11}. 
Common to these methods is that they provide a family of possible estimators instead of just one estimator, with the unregularized solution at one end of the spectrum. 
This family is indexed by a regularization parameter and is commonly referred to as the solution path. 
For the Lasso and ridge regression, this regularization parameter determines the extent of the respective penalties. For the iterative algorithms, this parameter corresponds to the number of steps they take. 
Despite the difference in nature, numerous works have shown these regularization methods, at least in the context of the linear model, are intrinsically related \citep{EfrHasJohTib04, ZhaYu07, MeiRocYu07}.
In that light, we will not focus on the distinction between the different types of regularization parameters but instead simply use $\lambda$ as a catch-all representation for them. 
In the same vein, we focus on the Lasso in this chapter even though we believe the method we present will work in the general framework.

\subsection{Selecting the Regularization Parameter $\lambda$}
Much work has been done to show that regularization methods yield desirable solutions in high dimensional problems.
For example, the popular Lasso has been shown to be $L_2$-consistent 
\citep{ZhaHua08,MeiYu06,BicRitTsy08} and model selection consistent \citep{MeiBuh06, ZhaYu06, Tro06, Wai09} in the high dimensional setting when respective conditions are met.
These results guarantee the existence of the $\lambda$ needed, but offer little guidance on how to find the desired $\lambda$ in practice.
Indeed, data-driven regularization parameter selection with guaranteed theoretical performance turns out to be a particularly difficult problem.

One can rely on traditional model selection criteria like Akaike's information criterion ($AIC$) \citep{Aka74} and Bayesian information criterion ($BIC$) \citep{Sch78}. 
They are easy to compute and have since been adapted for the high dimensional setting in the form of corrected $AIC$ \citep{HurTsa89} and extended $BIC$ \citep{ChenChen08}.
However, the validity of both the original and updated criteria rely on parametric assumptions.
Furthermore, they are derived from asymptotic results, so even when parametric assumptions are satisfied, they may not work well in the finite sample case. 

More commonly used today are parametric-model-free approaches like cross-validation ($CV$) \citep{All74, Sto74} and bootstrap methods \citep{Efr79, Zha93, Sha96}. Even though they too have asymptotic justifications, the heuristic rationale behind them are clear. Further, they have become computationally feasible for increasingly large data sets with the rapid advancements in computing power and the shift towards the parallel computipng paradigm.
These methods rely on data resampling to assess prediction error of candidate solutions and can be found in various statistics and machine learning literature \citep{HasTibFri02, Bre95, Bre96,Bre01}. 
In particular, it is the most popular approach used in regularization methods to select $\lambda$.
Doing so often leads to estimators with good predictive performance when the sample size is not small. 
However, there are other performance metrics that are also of interest in statistics, among them
parameter estimation and variable selection metrics, with important practical connections. 
Unsurprisingly, optimizing predictive performance does not necessarily translate to having success with respect to these other performance metrics.

\label{SecMotivation}
\subsection{Estimation Stability}
Statistical estimation is often tied to the optimization of an empirical loss 
or a random function based on data.
Take for example, when fitting a linear model for random variables $X \in \real^p, Y \in \real$, one might want to minimize the predictive $L_2$ loss, 
$$ f(\beta)= E_{X,Y}(Y-X'\beta)^2.$$
However, since the underlying joint distribution of $(X,Y)$ is unknown, we instead minimize the empirical loss  
$$ \hat f(\beta) = \frac{1}{n}\sum_{i=1}^n (y_i - x_i' \beta)^2,$$ where $(x_i,y_i)$ for $i=1,\ldots,n$,  are the observed samples of $(X,Y)$. By minimizing $\hat f$ instead of $f$, we incur a random estimation error dependent on the sample we observed. 
In the classical ideal scenario, when the sample contain independent and identically distributed observations and the sample size $n$ is large and
$p$ is small, 
this estimation error incurred is small. 
If we draw multiple samples from $(X,Y)$, each resulting estimate from minimizing the respective $\hat f$'s will be close to that of minimizing $f$, and consequently close to each other.
This closeness across different samples can be seen as a form of stability in the estimation procedure, and we call it \emph{estimation stability}. 

When the differences across different samples are measured by the $L_2$ error, the estimation stability is obviously related to variance. We opt to use
the term ``stability" rather than the more commonly used term ``variablity"
in statistics. This is to recognize the fact that stability is a concept broader than
variance or variability and that it is used in other quantative fields such as numerical analysis, dynamical systems, and linear analysis \citep{Hig96, LaS76, Ell98}.
Stability is also not associated with a particular metric (unlike variance)
and thus allows its consideration under different metrics.
In a recent paper \citep{Yu13}, we advocate
an enhanced emphasis on stability in statistical inference, especially
for large and high dimensional data for which instability of statistical methods is much more
common than in the domain of classical statistics.

It is clear that estimation stability is a necessary property for a reasonable estimation procedure: the solution is not meaningful if it varies considerably from sample to sample.
The converse certainly cannot be true in general: an arbitrary constant estimate will not vary but is certainly meaningless. 
Concurrent with and independent of our work, \citet{NanYan14} proposed diagnostic measures to investigate this instability.
For us, we make use of cross-validation, and devise a model-free criterion based on estimation stability for the selection of the regularization parameter $\lambda$.
Specifically, our proposed new criterion of estimation stability cross validation ($ESCV$)
combines a new metric of estimation stablity ($ES$) with $CV$.
For a given regularziation parameter $\lambda$,
our new $ES (\lambda)$  metric is the reciprocal of a test statistic 
for testing the null hypothesis that the regression function is zero. 
The test statistic is an estimate of the regression function standardized
by an approximate delete-d Jacknife standard error estimate based
on the same pseudo data sets as in $CV$, and both estimates are functions
of $\lambda$. The proposed $ESCV$ criterion chooses a local minimum
of $ES (\lambda)$ which is smaller (more regularized) than the selection of $\lambda$ by $CV$. It is worth noting that
the computational cost of $ESCV$ is similar to that of $CV$ and that they are
both well suited to parallel computation, the dominant computing platform for big data.

\subsection{Goal for $ESCV$}
We are focussed on the problem of selecting a regularization parameter $\lambda$, and the corresponding solution from the solution path. This is a practical problem faced by practitioners, who often turn to $CV$, and to a lesser extent, (extended) $BIC$. This may yield undesirable results depending on the circumstances and nature of the problem. For example, as shown in Section \ref{SecResults} the usual implementation of $CV$ has good predictive performance but poor model selection properties whereas $BIC$ works poorly in high noise situations.

We demonstrate that our criterion $ESCV$ provides a viable alternative to $CV$ and (extended) $BIC$.
We compare the three approaches with respect to several performance metrics when applied to the Lasso on both simulated data sets
with different predictor dependence set-ups and two real data sets. 
These performance metrics are $L_2$ error for parameter estimation,
prediction error, $F$-measure and model size for model selection performance.

To be clear, we acknowledge that it is unlikely for one solution in the solution path to be optimal on all fronts. However, we find that $ESCV$ is a strong candidate for a one solution compromise. 
We find that $ESCV$ compares favorably with $CV$ and $BIC$ where they are known to excel, and outperforms them in other scenarios over different performance criteria. 
In particular, $ESCV$ obtains excellent model selection results that
are substantially better than those from $CV$, both in simulations and our real data sets.
When the predictors are correlated, which is often the case in practice,
$ESCV$ also often outperforms $CV$ for parameter estimation 
while at same time provides prediction errors 
comparable to those of $CV$. 

We note that previous works based on stability of solutions have shown positive results in terms of model selection \citep{Bre96, Bac08, MeiBuh10}. 
The work here differs from them in three substantial ways. 
Firstly, we develop a different measure of stability $ES$
that is closely related to estimation rather than model selection, even though
our $ESCV$ does have desirable model selection properties quantified by the $F$-measure across all simulation set-ups in Section \ref{SecResults}.
Secondly, we restrict our attention to selecting the regularization parameter.  
Even though we evaluate our choice by the performance of the corresponding solution, our focus remains on determining the right amount of regularization.   
We do not introduce any further tuning parameters as in \citet{MeiBuh10}. 
Concurrent with and independent of our work, recent follow-up
papers on \citep{MeiBuh10} use model selection stability to select edges
in graphical models \citep{LiuRoeWas10,HauMorVerVer12} or modify stability model selection
to improve its false discovery rate theoretical properties \citep{ShaSam13}.
The former two papers introduce further tuning parameters and they
recommend fixed values for them.
\citet{ShaSam13} employs the complementary half-sample data perturbation scheme. $ESCV$ can work on such a scheme, but doing so would depart from the usual implementation of $CV$ for comparison purposes.
Thirdly, as in \citet{MeiBuh10}, these three papers apply 
data perturbation schemes such as bootstrap and subsampling with 
hundreds or thousands runs of model fitting. On the contrary,
the $CV$ (and $ESCV$) data perturbation
scheme often works well based on 5-10 runs of model fitting.

We also note the previous work on estimation stability in the computer science literature. \citet{BouEli02} defined algorithmic stability, and further works including \citet{KutNiy02, MukNiyPogRif06} explored the role stability has in some $M$-estimators. In particular, they show that good training stability is necessary and sufficient for consistency. The $ES$ metric we propose can be seen as a special form of some of the stability metrics in the above works. However, our goal is very different. We do not assume we have good training stability. Rather, we assert that that amongst all the candidate solutions, the $ES$ metric, can help select the best solution.

\section{Methodology}

\subsection{Lasso and Pseudo Solutions}
\label{LassoProblem}
Let $X \in \real^{n\times p}$, $Y \in \real^n$ be our data set. The Lasso generates a family of solutions,
$$\hat\beta[\lambda] = \arg\min_\beta \left\{||Y-X\beta||^2_2 + \lambda||\beta||_1\right\}.$$
$\hat\beta[\lambda]$, as a function of $\lambda\ge 0 $ is also known as the Lasso solution path for $\beta_j$ ($j=1,\ldots,p$). 
We want to select a solution from this solution path; that is, choose a $\lambda$ and take its corresponding solution in the solution path. 
As alluded to earlier, we would like to make this choice based on estimation stability and fit.

Since the notion of estimation stability is tied to the sampling distribution of the data, it is unavoidable that we need multiple solution paths to make such an assessment. Of course, it is often costly and infeasible to obtain extra data in practice. Thankfully, this problem is not new, and there are well-established ways to get around it. The key is to exploit the existing data by employing data perturbation schemes, parlaying it into multiple data sets. 
Let $(X^*[k], Y^*[k])$ represents our $k$th pseudo data set, derived from $(X,Y)$. In our case, these are the cross-validation folds:  we randomly partition the data into $V$ groups and form $V$ pseudo data sets by leaving out one group at a time. (See Section \ref{SecDis} for other data perturbation schemes.) We then get pseudo solutions,
$$\hat\beta[k;\lambda] = \arg\min_\beta \left\{||Y^*[k]-X^*[k]\beta||^2_2 + \lambda||\beta||_1\right\}$$
for $k = 1 \ldots V$.

\subsection{Alignment}
\label{SecAlign}
For many regularization methods, there are multiple representations for the regularization parameter $\lambda$. 
In the case of the Lasso above, $\lambda$ refers to the $L_1$ penalty parameter. Other popular choices to index the solution path are the $L_1$-norm of the coefficient estimate, and the $L_1$-norm expressed as a fraction of the $L_1$-norm of the unregularized solution. 
Each of these representations for the solution path has its own merits, and is equivalent to the others (when non-trivial) for any single solution path. The usual penalized least squares formulation of the Lasso as given in Section \ref{LassoProblem} is simply the Lagrangian form of the usual least squares problem subject to a constraint on the $L_1$-norm, and the $L_1$-norm of the unregularized solution is fixed for any single solution path.

\begin{figure}[h]
\begin{center}
\includegraphics[width=0.5\textwidth]{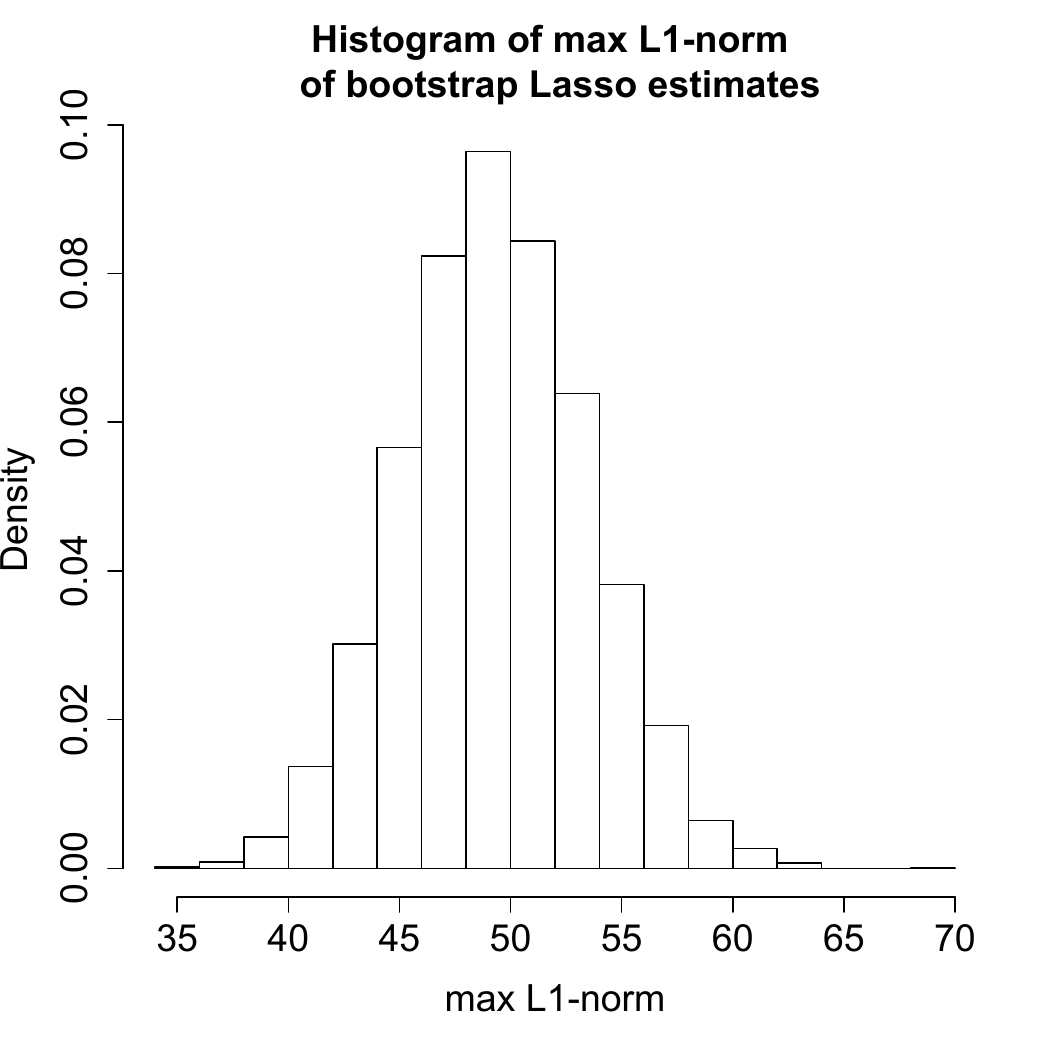}
\caption{Empirical bootstrap distribution of maximum $L_1$-norms of Lasso estimates on a typical simulated data set: a base case Gaussian simulation with $n=100$, $p=150$, $\sigma=1$, $\rho=0.5$ in Section \ref{SecCon}.}
\label{FigL1n}
\end{center}
\end{figure}

However, care must be taken on how to most meaningfully align our solution paths, when we reference the same $\lambda$ across different (pseudo) solution paths.
In particular, when $n <  p$, the $L_1$-norm of the unregularized solution corresponds to the saturated fit and can vary a lot depending on which data points were sampled. 
This makes $L_1$-fraction a poor choice, as the same index may correspond to very  different amounts of regularization.
The effect is more pronounced when the features are more correlated.   
Figure \ref{FigL1n} shows a histogram of the maximum $L_1$-norms for 10,000 bootstrap Lasso estimates of the base case Gaussian simulation (with $n=100$, $p=150$, $\sigma=1$, $\rho=0.5$) in Section \ref{SecCon}. There is considerable spread: in this case, the upper decile is over 20\% more than the lower decile.  

To highlight the effect of alignment on estimation performance, we compared the performance of cross-validation with the three alignments for the low noise scenarios detailed in Section \ref{SecCon}. As shown in Table \ref{AliTab}, aligning the solution paths with $L_1$-fraction does comparatively worse than aligning with $L_1$-norm or the penalty parameter. Notably, in the popular R package ``lars" used in solving the Lasso efficiently, the included cross-validation code aligns with $L_1$-fraction. 

For $ESCV$ to be proposed later, we find that there is little difference in performance when aligning with either the penalty parameter, $\lambda$ or the $L_1$-norm. In this work, we will be using the $\lambda$ alignment as it is seen as the canonical parameterization of the Lasso problem. This also allows us to make use of the increasingly popular R package ``glmnet'' \citep{FriHasTib10}, which can compute Lasso solutions considerably faster than competing methods.

\begin{table}[h]
\begin{center}
\begin{tabular}{|c|c|c|c|}
\cline{2-4}
\multicolumn{1}{c}{} & \multicolumn{3}{|c|}{Cross-Validation Estimation Error (Standard Error)} \\
\hline
$\rho$ & Regularization parameter &  $L_1$-norm & $L_1$-fraction\\
\hline
0 & 0.795 (0.005) & {0.792} (0.005) & 0.813 (0.005)  \\
0.2 & 0.788 (0.006) & {0.774} (0.005) & 0.827 (0.006)\\
0.5 & 0.967  (0.006) & {0.958} (0.006) & 1.03  (0.006) \\
0.9 & 1.83 (0.01) & {1.81} (0.01) & 1.93  (0.01) \\
\hline
\end{tabular}
\end{center}
\caption{Effect of alignment on cross validation performance on the base case Gaussian simulation with $n=100$, $p=150$, $\sigma=1$, in Section \ref{SecCon}.
The first column corresponds to the alignment based on $\lambda$,
the second based on $L_1$-norm and the third based on the $L_1$ fraction. Cross-Validation performs worst when aligning with $L_1$-fraction. The numbers are based on 1000 simulations.}
\label{AliTab}
\end{table}%

\subsection{Convergence of Pseudo Solutions}
Given $p$-dimensional pseudo solutions $\hat\beta[k;\lambda]$ for $k = 1,\ldots, V$, 
we want to measure their differences or see how similar or stable they are.
Computing their pair-wise $L_2$ errors
was a natural first attempt. 
However, we found that these errors vary too wildly to be useful
even after normalization by means when there is 
high dependence between the components in the vector and this happens
often especially when $p$ is large. 
Notice that the components of an estimate
of $\beta$ are combined in a linear fashion through $X\beta$
to achieve our primary goal of estimating the linear regression function.
Therefore we propose to compute the estimates 

$$\hat Y[k;\lambda] = X \hat\beta[k;\lambda],$$

\noindent
and study their stability.

To evaluate such stability, as mentioned earlier we need a measure for how far apart the estimates are at each $\lambda$: stable pseudo solutions should give similar estimates. One possibility is to look at the average pairwise squared Euclidean distance between the $V$ estimates: $$A(\lambda) := \frac{1}{{V \choose 2}} \sum_{k \neq j} ||\estim{Y}[k; \lambda] - \estim{Y}[j; \lambda]||^2_2.$$
It is not hard to see that this is proportional to the more familiar ``sample variance'' formulation,
$$\estim\Var(\hat Y[\lambda]) = \frac{1}{V} \sum_{k=1}^V ||\hat Y[k;\lambda] - \bar{\hat{Y}}[\lambda] ||^2_2,$$
where $ \bar{\hat{Y}}[\lambda] = \frac{1}{V} \sum_{i=1}^V \hat Y[i;\lambda]$.

\begin{figure}[h]
\begin{center}
\includegraphics[width=0.85\textwidth]{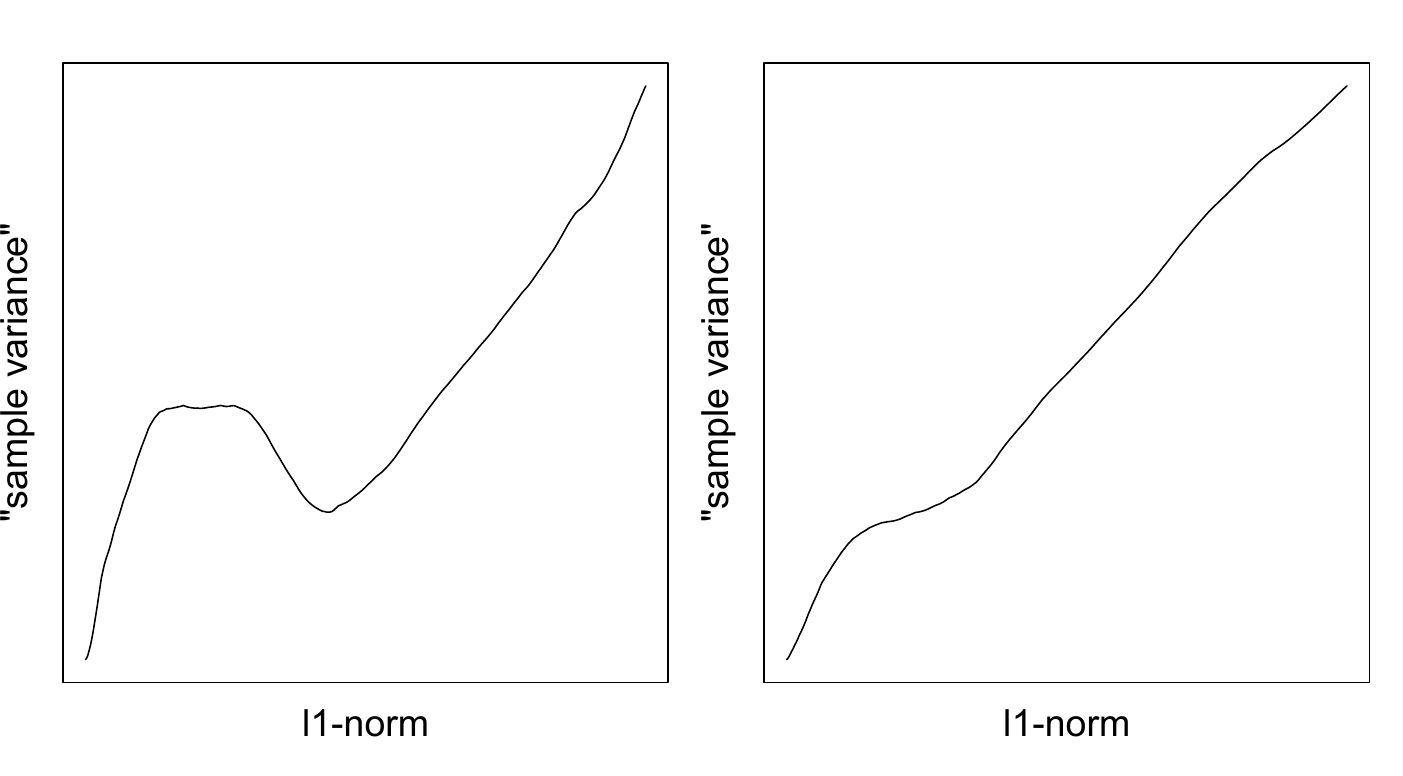}
\caption{Examples of the sample variance metric. The left panel shows an example where the metric exhibits a ``dip'', representing the ``convergence'' of the pseudo solutions. The right panel shows an example with a much muted ``dip''. It is difficult to use the sample variance metric to select a solution on the right.}
\label{FigSS1}
\end{center}
\end{figure}

Figure \ref{FigSS1} shows two examples of this sample variance metric. Here, the metric is indexed by $L_1$-norm of the original solution path for better visualization. The left panel is particularly illuminating: the pseudo solutions diverge as they grow at first but converge somewhat before diverging again.
Here, \emph{convergence} and \emph{divergence} simply refer to the sample variance metric  (which is really just the average pairwise distance) decreasing and increasing respectively.
Heuristically, this behavior is exactly what one would expect if there is a ``correct'' amount of regularization. Different samples would take different paths towards the ``correct'' solution before moving away from one another due to overfitting. 
Hence, we might select the $\lambda$ corresponding to the minimum point \emph{after} the first negative slope.
That is, we want to choose $\lambda$ corresponding to the ``dip''.

By doing this, we incorporate fit into our selection even though our criterion is based on stability. 
The \emph{convergence} of the solution paths is key: not only does it suggest we are close to the truth, we are also gifted
with estimation stability.
Note that this helps us automatically exclude $\lambda$'s where the solution paths trivially agree. We see this trivial effect in Figure \ref{FigSS1}, where the global minimum for the sample variance metric occurs where the solutions are close to zero. 

However, this convergence effect is not always clear. The ``dip'' is not always present as shown in the example on the right panel. There you can still see the drop in gradient, but it is not clear which $\lambda$ we should pick. Notice, however, that in a solution path, the norm of the solution varies with the amount of regularization (by definition in our case). Since larger solutions naturally varies more, using the sample variance metric skews the choice towards solutions with small norms. We need to bring in the concept of normalization to account for this effect.

\subsection{Hypothesis Testing and the Estimation Stability Metric}
In hypothesis testing, a test statistic based on data is computed and its corresponding $p$-value is calculated by matching the test statistic with its model-specific theoretical distribution. This test statistic often takes the form of a mean value over its estimated standard deviation, e.g. the student's $t$-test. The desired outcome for the $t$-test, as is often the case regardless of the assumed model and $p$-value computation, is to have the test-statistic away from 0. The heuristic there is clear: if the hypothesized effect is real, the size of the mean value should be large compared to its estimated standard deviation.

\begin{figure}[h]
\begin{center}
\includegraphics[width=0.85\textwidth]{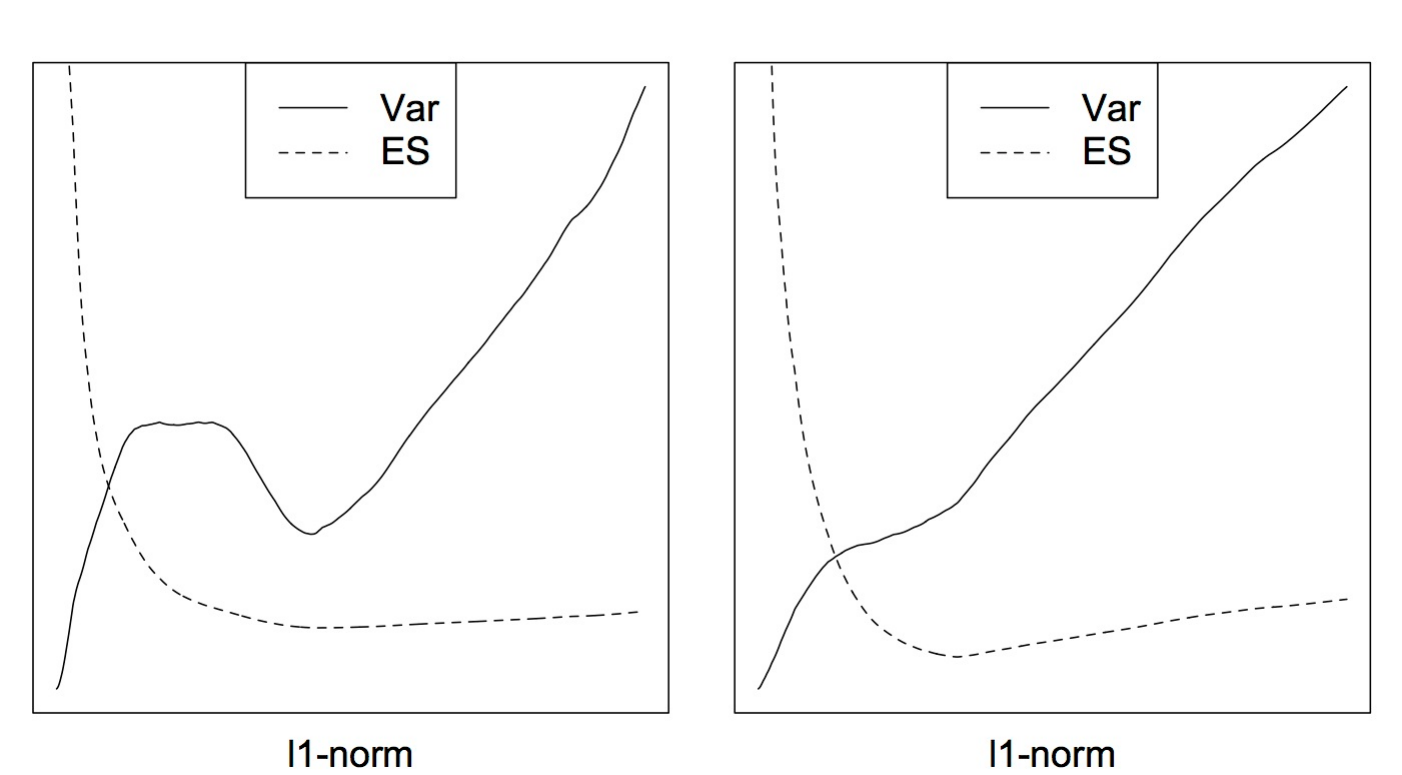}
\caption{Examples of the sample variance metric (as in Figure \ref{FigSS1}) and the corresponding $ES$ metric. We see that the $ES$ metric preserves the local minimum from the sample variance metric on the left panel, and introduces one on the right panel where there was no local minimum from the sample variance metric.}
\label{FigSS2}
\end{center}
\end{figure}

In the same vein, our sample variance metric should be relative to the squared mean size of the corresponding solution. We define the \emph{estimation stability metric}, 
$$ES(\lambda) := \frac{\estim\Var(\hat Y[\lambda])}{||\bar{\hat{Y}}[\lambda]||^2_2},$$
the normalized version of the sample variance metric. Figure \ref{FigSS2} shows the corresponding $ES$ metrics in dashed lines superimposed on the old sample variance metric. On the left, the ``dip'' from the sample variance metric is preserved by the $ES$ metric. On the right, there is now a pronounced minimum we can select.

A related instability measure is defined in \citet{YuaYan05}. It is a function of the size of the data perturbation, and is not normalized by the solution size as in $ESCV$, but instead by an estimate of the noise in the model. This is applied in the context of a small number of models to be used in model averaging. In our case, we have a large number of candidate $\lambda$'s, and our goal is to find one best solution in the solution path.

The $ES$ metric's reciprocal has exactly the form of a test-statistic. We can view the $ES$ selection of $\lambda$ as a set of hypothesis tests. For each $\lambda$, we are testing if the fit ($\hat Y[\lambda]$) is statistically different from fitting the null model ($E(Y) = 0$), albeit without a specified theoretical distribution. Our $ES$ criterion of choosing the $\lambda$ corresponding to the convergence of pseudo solutions, is exactly choosing $\hat Y[\lambda]$ with locally minimal normalized variance. This in turn, is exactly choosing the solution whose $ES$ metric has the largest reciprocal, or in our analogy, the most statistically significant solution along the path.

\subsection{$ESCV$: Incorporating Cross-Validation}
\label{SecIncCV}
There is no guarantee that our $ES$ metric would have only one local minimum. Unless the multiple solution paths match up perfectly, there will be a local minimum or multiple local minima. 
Hence, even in the case where $Y$ bears no relation to $X$ at all, an inadvertent minimum on the $ES$ metric will falsely suggest the pseudo solutions are converging towards a meaningful solution.
To prevent scenarios like this where $ES$ fails, we incorporate cross-validation into our selection.
We have already limited our choice of minimum $ES$ to local minima. Here we further limit it to the local minimum of $\lambda$ that gives a smaller solution than the cross-validation choice. We call this improved criterion \emph{estimation stability with cross validation} ($ESCV$).
In Section \ref{SecResults} on experimental results, we use
a grid-search algorithm to find such a local minimum of $ES$ as
commonly done for $CV$. Thus $ESCV$'s computational cost is similar
to that of $CV$ and they are both easily parallelizable.

We are exploiting the fact that cross-validation overselects \citep{LenLinWab06, WasRoe09}. (Please see Section \ref{SecCVc} for more details.) When $ES$ gives a meaningful local minimum, cross-validation will likely overselect. Hence, $ESCV$ behaves like $ES$ above. 
However, when $Y$ bears no relation to $X$, or when the noise overwhelms the signal, cross-validation will likely choose the trivial solution correctly. In this case, $ESCV$ will follow suit and pick up the trivial solution.
Note that this has negligible additional computation cost, as we are essentially getting the cross-validation choice for free. The bulk of the computation lies in computing the multiple solution paths we already have. 

\subsection{$ESCV$ the Method}

To sum up, we have devised a $ES$ metric which measures estimation stability. 
$$
ES(\lambda) := \frac{\frac{1}{V} \sum_{k=1}^V ||\hat Y[k;\lambda] - \bar{\hat{Y}}[\lambda] ||^2_2}{||\bar{\hat{Y}}[\lambda]||^2_2},
$$
where $ \bar{\hat{Y}}[\lambda] = \frac{1}{V} \sum_{i=1}^V \hat Y[i;\lambda]$.

We would like to select a $\lambda$ that minimizes $ES(\lambda)$, but at the same time encompass the convergence effect of pseudo solutions as well as leverage the $CV$ choice for fit information. Our choice $\lambda_{ESCV}$ is a local minimum of $ES(\lambda)$ that gives a smaller solution than the $CV$ choice. That is,
$$
\lambda_{ESCV} = \arg\min_{\lambda \in \Lambda} ES(\lambda) , 
$$
where
$$\Lambda = \left\{ \lambda \ge \lambda_{CV}  \ \middle| \ 
ES(\lambda) = \min_{\omega \in (\lambda - \epsilon, \lambda + \epsilon)} ES(\omega)
\text{ , for some $\epsilon > 0$.}\right\}
$$
Note that $\lambda_{ESCV} \ge \lambda_{CV}$  is equivalent to $||\hat\beta[\lambda_{ESCV}]||_1 \le  ||\hat\beta[\lambda_{CV}]||_1$). If there exist multiple local minima, our choice corresponds to the minimal value of $ES(\lambda)$ amongst the local minima. In the rare case where there is no local minima ($\Lambda = \emptyset$), we drop the condition and simply choose $\lambda_{ESCV} = \arg\min_{\lambda \ge \lambda_{CV}} ES(\lambda)$.

Our method assumes there is no intercept term in the linear model. If this is not a reasonable assumption, we should first center the data.

\subsection{Discussion on $ESCV$}
\label{SecDis}

Our $ES$ metric is based on assessing the stability of the fitted values $\hat Y[\lambda] = X \hat\beta[\lambda]$ instead of the estimates $\hat\beta[\lambda]$. 
This seems counter-intertuitive since we are interested in a variety of performance measures, most of which are based on the quality of $\hat\beta[\lambda]$ itself.
However, we note that these performance measures only make sense if the underlying $\beta$ is identifiable.
To that end, there is a large volume of work showing the Lasso is model selection consistent under regularity conditions including that the smallest
non-zero true parameter value is not too small compared to a rate decaying in $n$ \citep{MeiBuh06, Tro06,ZhaYu06,  Wai09}. In particular, it assures us the asymptotic recovery of the underlying true $\beta$ under appropriate conditions.

However, in the finite sample case, and especially when the features are highly correlated, different linear combinations of features (of a given sparsity) may give approximately equivalent fits. Under data perturbation, it is not surprising that the different solution paths choose different features. This makes any metric based on $\hat\beta[\lambda]$ statistically unstable since $V$ is small. Note that this does not contradict the assessment of the eventual $\hat\beta[\lambda]$ picked since $ESCV$ and $CV$, picking from the same solution path, would both suffer from any failure of the original Lasso. 

In $ESCV$, we have used cross-validation folds to compute our pseudo-solutions.
There are of course many other ways to generate pseudo datasets. 
One related approach would be to apply bootstrap sampling~\citep{Bac08}. 
Here, simply sample with replacement from the original data set to generate multiple data sets.
These two approaches are obvious choices, and can be applied to any estimation procedure (even those without an optimization formulation).  A third choice, which applies only to penalized $M$-estimators such as the Lasso, is based on perturbations of the penalty~\citep{MeiBuh10}.
Note that such perturbations of the penalty amount to perturbing (indirectly) the samples, but in a different way than bootstrapping.
Finally, we can simply perturb the data directly by adding noise to $X$ and/or $Y$. 
For example, we can add random Gaussian noise to the response \citep{Bre96}. 
We find in our experimental results that the choice of data perturbation scheme (within reason) does not change our narrative of how $ESCV$ behaves. The same convergence effect is observed, and the resulting $ESCV$ pick is reasonable in terms of the performance metrics. 

With high dimensional data, computation can be costly. In the case of the Lasso, 
 even with efficient algorithms, the computation quickly gets expensive with larger data sets
\citep{EfrHasJohTib04,MaiYu12}. Using the estimation stability metric to select the regularization parameter incurs only as much computation as using cross validation. This is because the bulk of the computation in both cases rests in computing the solution paths of the $V$ perturbed data sets. $V$ in this case can be small as demonstrated in Section \ref{SecResults}.
This is in contrast to related work \citep{Bac08, MeiBuh10} which requires a much larger $V$.

\subsection{Discussion on Choice of $V$ in $CV$}
\label{SecCVc}

\citet{ArlLer12} investigated the effect of $V$ on $CV$ performance. They found that the variance of the solution decreases as you increase $V$ but asymptotes quickly. This coincides with the conventional wisdom of choosing $V = 10$. In our experience with $ESCV$, perhaps unsurprisingly given we are using the same pseudo data sets, we have found the same effect when varying $V$ from 2 to 20. Note that this variance reduction is of the final solution, not the fits of the pseudo data sets.

\citet{Sha93} was motivated by the model inconsistency of leave-one-out cross validation. He showed that this can be rectified by using a validation set of size $n_v$, satisfying $n_v/n \rightarrow 1$. Note that this condition is not met for any fixed choice of $V$. We refer the reader to \citet{Yan07} for more on the data splitting ratio. As pointed out in \citet{Yan07}, \citet{Zha93} showed that $V$-fold $CV$, amongst other variations of $CV$, is inconsistent for any fixed splitting ratio. Nevertheless, these works suggest that a smaller $V$ for $CV$ will result in better model selection. We find this to be true in our simulations; $CV$ overselects less with a smaller $V$, but overselects nonetheless.

For all our results in Section \ref{SecResults}, we will present the results with the conventional choice of $V = 10$ for both $ESCV$ and $CV$. We also compare $CV$ with different choices of $V$ along with $ESCV$ in our fMRI example in Section \ref{SecfMRI2}, as it offers an unique opportunity where the predictive performance is similar over a large range of model size. 

\section{Experimental Results for Lasso}
\label{SecResults}
In this section, we evaluate $ESCV$'s performance relative to the cross validation ($CV$) across a variety of data examples. In each problem, we fit a linear model using the Lasso. We focus our attention on the comparison with $CV$ as it is the most popular criterion in practice. The R code for all the simulations is included as supplementary material.

In all the data examples, we use the same grid-search algorithm
to evaluate our $ES$ and $CV$ metrics. 
For our algorithm, we first run Lasso on the original data using the R package ``glmnet'', which determines the grid of 100 candidate $\lambda$'s. As documented in \citet{FriHasTib10}, the grid starts with the smallest $\lambda_{max}$ that gives $||\hat\beta[\lambda_{max}]||_1 = 0$, and decreases uniformly on a log scale. The minimum $\lambda$ on the grid depends on the relationship of $n$ and $p$. The $\lambda$ grid is then used on all pseudo data sets to evaluate our $ES$ and $CV$ metrics.

We start with simple sparse gaussian linear model simulations with our focus on the high dimensional data set up. We will vary the simulation parameters such as correlation strength within features and signal strength, as well as explore popular correlation structures of the design matrix, to cover a wide 
range of data scenarios in practice. We compare the solutions picked by $ESCV$ and $CV$ with regard to parameter estimation, prediction, and model selection performance measures such as $F$-measure and model size. We also include the extended $BIC$ choice, and follow the suggestions by the authors \citep{ChenChen08} on the choice of its tuning parameter $\gamma$. For most of the simulations, we use $\gamma = 0.5$ as they fall under the high dimensional setting. The only exception is the $n=100,\ p=50$ case, where we use the original $BIC$, corresponding to $\gamma = 0$.

We also explore the performance of our method on two real data sets from neuroscience and bioinformatics. We use a combination of objective predictive performance and subject knowledge on plausible models to illustrate the efficacy of $ESCV$ over $CV$. In all cases, note that we are comparing different choices of $\lambda$ on the same solution path (from the original data). Furthermore, we use the same data splits to make comparable results of $CV$ and $ESCV$. 

\subsection{Gaussian Simulation}
Let $X_i \in \real^p $ for $i = 1,\ldots,n$ be independent identically distributed Gaussian variables with mean 0 and covariance $\Sigma$. 
We have the usual linear model
$ Y_i = X_i'\beta + \epsilon_i,$
where $\beta \in \real^p$ is the unknown parameter, and $\epsilon_i \in \real$ is independent Gaussian noise with standard deviation $\sigma$. 
$\beta_j$ are drawn from $U[\frac{1}{3},1]$ for $j = 1, \dots, 10$ and 0 otherwise.
The separation from zero is for model selection to make sense. 
This is a common assumption in theoretical work.
We have found that other patterns of coefficients behave similarly as long as the smallest coefficient is well-separated from 0 relative to the average coefficient size.

The reported estimation and prediction errors are defined as $$||\hat\beta-\beta||_2 \textrm{ and } \sqrt{E_X(||X\hat\beta - X\beta||_2^2)}=\sqrt{(\hat\beta - \beta)'\Sigma(\hat\beta - \beta)}$$ respectively.
For model selection, we use the $F$-measure which balances false positive and false negative rates of identifying non-zero coefficients of $\beta$. The higher the $F$-measure the better.
Each simulation is repeated 1000 times and the performance measures are aggregated across them.

\subsubsection{A Base Case}
\label{SecCon}
Within the Gaussian linear model setup, there are many problem scenarios that favor one method over others.
In particular, the following problem settings are known to affect the performance of the Lasso: correlation strength between features, strength of signal (size of coefficients) relative to the noise levels, dimension of the problem ($p$), and the correlation structure of the features.
This is of course not an exhaustive list but is sufficient to cover a wide range of problems.
As the strength of the correlation and signal are key to the behavior of the Lasso solution, we will include a full complement of these problem settings to illustrate when and why $ESCV$ works well.

We start with a base case scenario.
Here, $\Sigma$ has entries 1 down the diagonal and constant $\rho$ on the off-diagonal. 
We vary $\rho = 0, 0.2, 0.5, 0.9$ and $\sigma = 0.5,1,2.$
We set $n=100$ and $p=300$ to emulate the high dimensional data setting.
Note that this implies that the columns of $X$ are empirically correlated even when the features they represent are independent.

As expected, $CV$ does well in terms of prediction error (see Table \ref{ConTab}). However, observe that this does not necessarily translate to success in terms of other performance measures.
With estimation error, we find that once we leave the orthogonal case $\rho=0$ where estimation and prediction error are equivalent, $ESCV$ has lower estimation error than $CV$ despite having comparable prediction error. 

For model selection, we use the $F$-measure, the harmonic mean of the precision and recall rates, which are inversely proportional to false positive rate and false negative rate respectively. A high $F$-measure is achieved when both false positive and false negative rates are low.
Recall that we are selecting solutions from the same solution path. The Lasso solution path corresponds roughly to a nested family of models in terms of features picked since features seldom gets dropped as we relax the penalty term. Hence, having a low false negative rate (high recall) typically comes at the cost of a high false positive rate (low precision). The $F$-measure balances these two objectives. 

By this measure, $ESCV$ often outscores $CV$ by a considerable margin. $CV$ picks more true variables, but in the process picks up a disproportionately large number of noise variables. This is in line with theory that $CV$ often overselects~\citep{WasRoe09}. $ESCV$ cuts down the false positive rate, but not too much at the expense of the false negative rate.

The extended $BIC$, designed to achieve model consistency, does well in terms of model selection, but poorly in estimation and prediction. It does exceptionally well in the low noise setting, but progressively worse as we increase the noise. This is not unexpected since $BIC$'s model selection consistency is an asymptotic result, and high noise levels can be seen as the non-asymptotic case.
Comparatively, $ESCV$ maintains its good model selection performance and overtakes $BIC$ in the higher noise settings.

The results are summarized in Table \ref{ConTab} and the standard errors (SE) are given in Table \ref{ConTabSE}.
Note that the performance measures are highly correlated since for each simulation run, the selections by $ESCV$, $CV$ and $BIC$ are from the same solution path. Hence, the SEs for paired differences in performance measures are actually lower than the SEs for each of the values as reported in Table \ref{ConTabSE}. 

\begin{table}[htbp]
\setlength{\tabcolsep}{3pt}
\begin{center}
\resizebox{.8\textwidth}{!}{%
\begin{tabular}{|c|c|c|c|c|c|c|c|c|c|c|c|c|c|}
\cline{3-14}
\multicolumn{2}{c}{} & \multicolumn{3}{|c|}{Estimation} & \multicolumn{3}{c|}{Prediction} & \multicolumn{3}{c|}{Model Selection} & \multicolumn{3}{c|}{Model Size}\\
\multicolumn{2}{c}{} & \multicolumn{3}{|c|}{error} & \multicolumn{3}{c|}{error} & \multicolumn{3}{c|}{$F$-measure} & \multicolumn{3}{c|}{} \\
\hline
$\rho$ & $\sigma$ & $ESCV$ & $CV$ & $BIC$ & $ESCV$ & $CV$ & $BIC$& $ESCV$ & $CV$& $BIC$ & $ESCV$ & $CV$& $BIC$\\
\hline
0 & 0.5 & 0.536 & \bf{0.471} & 0.629 & 0.536 & \bf{0.471} & 0.629 & 0.579 & 0.351 & \bf{0.726} & 24.4 & 47.0 & 17.2 \\
0 & 1 & 1.03 & \bf{0.934} & 1.56 & 1.03 & \bf{0.934} & 1.56 & 0.496 & 0.341 & \bf{0.501} & 26.3 & 46.9 & 9.91\\
0 & 2 & 1.69 & \bf{1.65} & 2.15 & 1.69 & \bf{1.65} & 2.15 & \bf{0.356} & 0.332 & 0.0646 & 24.1 & 31.4 & 1.57\\
\hline
0.2 & 0.5 & \bf{0.484} & \bf{0.484} & 0.508 & 0.480 & \bf{0.471} & 0.523 & 0.479 & 0.418 & \bf{0.523} & 31.7 & 37.7 & 28.1\\
0.2 & 1 & \bf{0.872} & 0.886 & 1.02 & 0.822 & \bf{0.816} & 1.14 & 0.447 & 0.379 & \bf{0.506} & 33.7 & 41.7 & 24.6\\
0.2 & 2 & \bf{1.56} & 1.61 & 2.04 & \bf{1.48} & 1.49 & 3.17 & \bf{0.381} & 0.329 & 0.216 & 30.2 & 38.2 & 2.88\\
\hline
0.5 & 0.5 & \bf{0.679} & \bf{0.679} & 0.700 & 0.584 & \bf{0.582} & 0.617 & 0.444 & 0.429 & \bf{0.469} & 34.4 & 35.9 & 31.9\\
0.5 & 1 & \bf{1.10} & 1.12 & 1.15 & \bf{0.824} & 0.830 & 0.933 & 0.413 & 0.375 & \bf{0.441} & 35.5 & 40.3 & 31.0\\
0.5 & 2 & \bf{1.78} & 1.85 & 1.93 & \bf{1.32} & 1.35 & 2.41 & \bf{0.338} & 0.302 & 0.330 & 30.9 & 36.8 & 16.0\\
\hline
0.9 & 0.5 & \bf{1.53} & \bf{1.53} & 1.58 & 0.722 & \bf{0.721} & 0.778 & 0.363 & 0.363 & \bf{0.368} & 33.0 & 33.0 & 30.2\\
0.9 & 1 & 1.98 & \bf{1.97} & 2.04 & 0.733 & \bf{0.722} & 0.850 & \bf{0.297} & \bf{0.297} & 0.294 & 29.6 & 30.2 & 25.2\\
0.9 & 2 & \bf{2.56} & 2.66 & 2.59 & \bf{0.880} & 0.882 & 1.18 & \bf{0.186} & 0.179 & 0.173 & 20.8 & 24.1 & 15.9\\
\hline
\end{tabular}}
\end{center}
\caption{Performance of $ESCV$, $CV$ and extended $BIC$ in picking the regularization parameter for the Lasso for our base case design: constant correlation $\rho$, $n=100$, $p = 300$. We see that $ESCV$ performs best in parameter estimation (when different from prediction) and model selection, while doing comparably to $CV$ in prediction. }
\label{ConTab}
\end{table}%

\begin{table}[htbp]
\begin{center}
\resizebox{0.8\textwidth}{!}{%
\begin{tabular}{|c|c|c|c|c|c|c|c|c|c|c|}
\cline{3-11}
\multicolumn{2}{c}{} & \multicolumn{3}{|c|}{Estimation} & \multicolumn{3}{c|}{Prediction} & \multicolumn{3}{c|}{Model Selection} \\
\multicolumn{2}{c}{} & \multicolumn{3}{|c|}{error SE} & \multicolumn{3}{c|}{error SE} & \multicolumn{3}{c|}{$F$-measure SE}  \\
\hline
$\rho$ & $\sigma$ & $ESCV$ & $CV$ & $BIC$ & $ESCV$ & $CV$ & $BIC$& $ESCV$ & $CV$& $BIC$ \\
\hline
0 & 0.5 & 0.003 & 0.003 & 0.005 & 0.003 & 0.003 & 0.005 & 0.004 & 0.003 & 0.006\\
0 & 1 & 0.008 & 0.006 & 0.02 & 0.008 & 0.006 & 0.02 & 0.005 & 0.003 & 0.01\\
0 & 2 & 0.008 & 0.008 & 0.007 & 0.008 & 0.008 & 0.007 & 0.004 & 0.004 & 0.005\\
\hline
0.2 & 0.5 & 0.004 & 0.004 & 0.004 & 0.005 & 0.005 & 0.005 & 0.002 & 0.003 & 0.003\\
0.2 & 1 & 0.005 & 0.005 & 0.01 & 0.005 & 0.005 & 0.02 & 0.002 & 0.003 & 0.004\\
0.2 & 2 & 0.007 & 0.008 & 0.01 & 0.008 & 0.007 & 0.02 & 0.003 & 0.003 & 0.009\\
\hline
0.5 & 0.5 & 0.008 & 0.008 & 0.008 & 0.009 & 0.009 & 0.01 & 0.002 & 0.002 & 0.002\\
0.5 & 1 & 0.007 & 0.007 & 0.008 & 0.006 & 0.006 & 0.01 & 0.002 & 0.002 & 0.002\\
0.5 & 2 & 0.008 & 0.009 & 0.009 & 0.006 & 0.006 & 0.04 & 0.002 & 0.003 & 0.004\\
\hline
0.9 & 0.5 & 0.01 & 0.01 & 0.01 & 0.01 & 0.01 & 0.01 & 0.003 & 0.003 & 0.003\\
0.9 & 1 & 0.009 & 0.009 & 0.01 & 0.008 & 0.008 & 0.01 & 0.003 & 0.003 & 0.003\\
0.9 & 2 & 0.009 & 0.01 & 0.009 & 0.005 & 0.004 & 0.01 & 0.003 & 0.003 & 0.003\\
\hline
\end{tabular}}
\end{center}
\caption{Standard errors (SE) for performance numbers in Table \ref{ConTab}.}
\label{ConTabSE}
\end{table}%

\subsubsection{Effect Of Ambient Dimension}
We repeat the simulations but this time for $p=50$ and $p=500$ to investigate the effect of the ambient dimension.
Note that only the number of non-relevant features is changing; the number of non-zero coefficients remain at 10, the sample size $n$ remains at 100.
The comparison of $ESCV$ and $CV$ from the base case extends here:
$CV$ does well in prediction error, especially in the independent predictors case, but loses out to $ESCV$ in the other scenarios
with dependence more relevant to practice and in terms of
parameter estimation and model selection metrics that are important
for scientific applications. The results are summarized in Table \ref{Amb1Tab} and \ref{Amb2Tab}.

As noted above, we use the original $BIC$ for the $p=50$ case and the extended $BIC$ for $p=500$. Again, the results from the base case extends. $BIC$ does well in terms of model selection, but its performance drops off quickly across all performance metrics as the noise level increases. 

\subsubsection{Other Correlation Structures}
The constant correlation structure can be seen as a simple one latent variable model.
Here we introduce other correlation structures corresponding to more complex models and run the same simulations ($n=100, p=300$, and varying $\sigma$ and $\rho$).
First, block correlation: all $p$ features are randomly grouped into 10 blocks, and within each block, the features have correlation $\rho$ while features from separate blocks are independent. Here, we let $\rho = 0.3, 0.5, 0.9$. 
Second, Toeplitz design: $\Sigma_{ij} = \rho^{|i-j|}$, with $\rho = 0.5, 0.9, 0.99$. In both cases, the ten true variables indices are randomly distributed among the $p$ variables so that they are not all strongly correlated with each other. The results for the two designs are summarized in Tables \ref{BloTab} and \ref{ToeTab} respectively.

Despite the different correlation structures, the qualitative results from the prior section holds again in both variations. For prediction error, $CV$ almost always outperforms $ESCV$, but $ESCV$'s predictive performance can be quite close to $CV$'s when $\rho \neq 0$. For estimation error, $ESCV$ gains on and eventually outperforms $CV$ with increasing correlation levels. And for model selection, $ESCV$ almost always has a higher $F$-measure than $CV$. Digging deeper, Table \ref{TPFPTab} shows the breakdown of the $F$-measure into the true positive and false positive rates. We can see that $ESCV$ has much lower false positive rates while sacrificing relatively little on the true positive rates.

\subsection{fMRI Data}
\label{SecfMRI}
This data is from the Gallant Neuroscience Lab at University of California, Berkeley. In this experiment, a subject is shown a series of randomly selected natural images and the fMRI response from his primary visual cortex is recorded. The fMRI response is recorded at the voxel level, where each voxel corresponds to a tiny volume of the visual cortex. The task is to model each voxel's response to the $n=1500$ images. The image features are approximately ${10000}$ transformed Gabor wavelet coefficients. We evaluate the prediction performance by looking at correlation scores against an untouched validation set of 120 images with 10-13 replicates. 
There are 1250 voxels in all. We ranked them according to their predictive performance under a different procedure from a previous study \citep{KayNasPreGal08}. Not all of them are informative, so we only look at the top 500.

\begin{figure}[h]
\begin{center}
\includegraphics[width=0.55\textwidth]{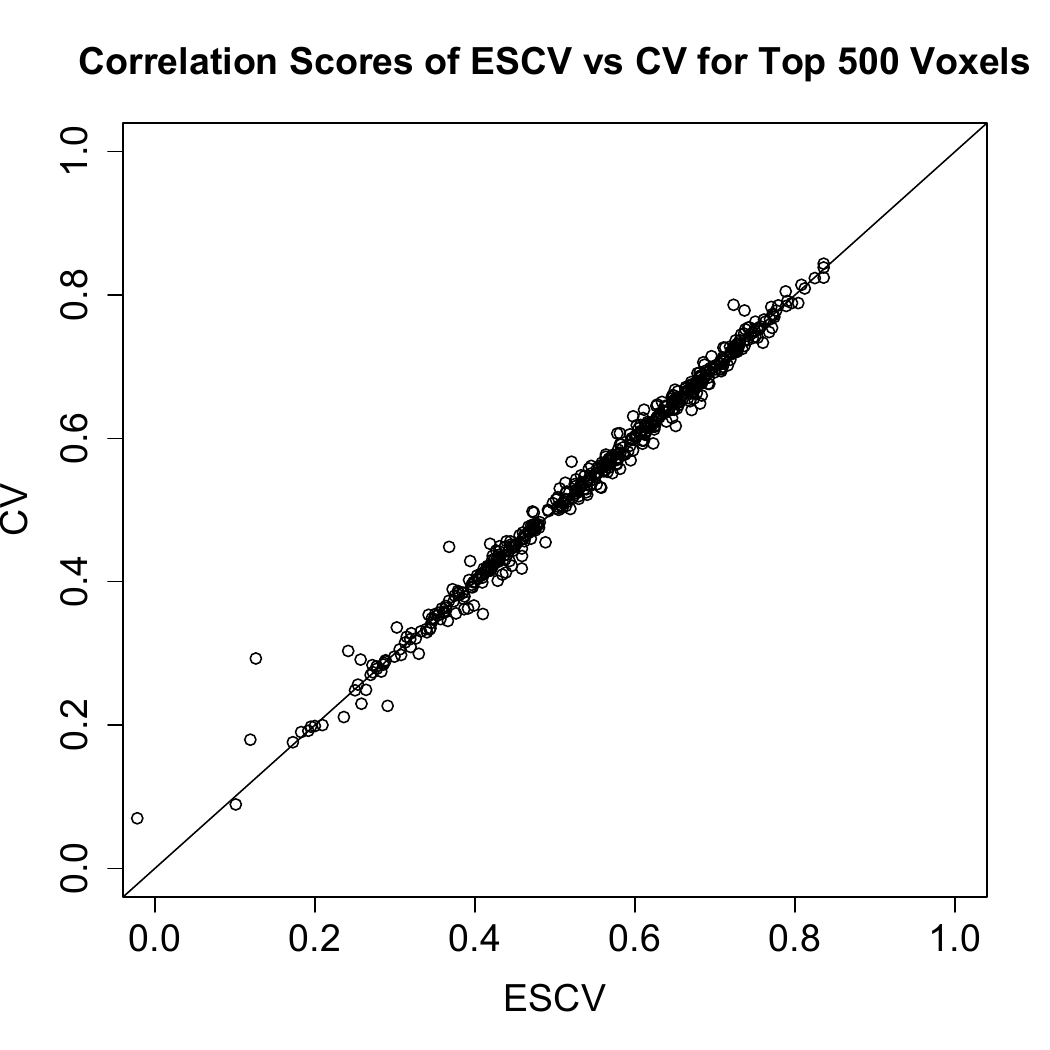}
\caption{Scatterplot of predictive correlation scores of $ESCV$ and $CV$ for the top 500 voxels in the fMRI data set. We see that for almost all 500 voxels, the predictive performances are similar for $ESCV$ and $CV$.}
\label{FigCors}
\end{center}
\end{figure}

We find that while the prediction performance are nearly identical for $ESCV$ and $CV$, $ESCV$ selects much fewer features. The results are in Table \ref{fMRITab}. For the sake of brevity, they are averaged across groups of 100 voxels. For example, for the top 100 voxels, on average, the correlation scores are similar, but $ESCV$ selects 30 features compared to $CV$'s 70 features - a close to 60\% reduction. That is, $ESCV$ selects a much simpler and also more reliable model that predicts just as well as $CV$. Figure \ref{FigCors} shows how close the correlation scores are.

\begin{table}[htbp]
\begin{center}
\begin{tabular}{|c|c|c|c|c|}
\hline
Voxels  & \multicolumn{2}{c}{Correlation Score} & \multicolumn{2}{|c|}{Model Size}  \\
\cline{2-5}
 & $ESCV$ & $CV$ & $ESCV$ & $CV$ \\
\hline
1-100 & 0.730 & 0.735 & 30.1 & 70.2 \\
101-200 & 0.653 & 0.655 & 27.0 & 61.8 \\
201-300 & 0.567 & 0.566 & 22.6 & 49.6 \\
301-400 & 0.455 & 0.459 & 16.7 & 40.3 \\
401-500 & 0.347 & 0.347 & 16.5 & 33.6 \\
\hline 
\end{tabular}
\end{center}
\caption{Performance on fMRI data set. The numbers are averaged across the respective hundred voxels. $ESCV$ cuts down the model size by more than half compared to $CV$, while largely preserving prediction accuracy.}
\label{fMRITab}
\end{table}%

We note again that $ESCV$ picks fewer features than $CV$ by design (Section \ref{SecIncCV}). That being said, the reduction is huge here: $ESCV$ picks less than \emph{half} the number of features as $CV$ across the different voxels. 
Furthermore, this was with little or no loss in predictive performance. To understand the results better, we look at the individual voxels and examine the features selected. In almost all the cases, $ESCV$ selects a subset of the features selected by $CV$. This is because they both select from the same Lasso solution path and features are rarely dropped after being added to the solution as we relax the regularization. 

\begin{figure}[htbp]
\begin{center}
\includegraphics[width=\textwidth]{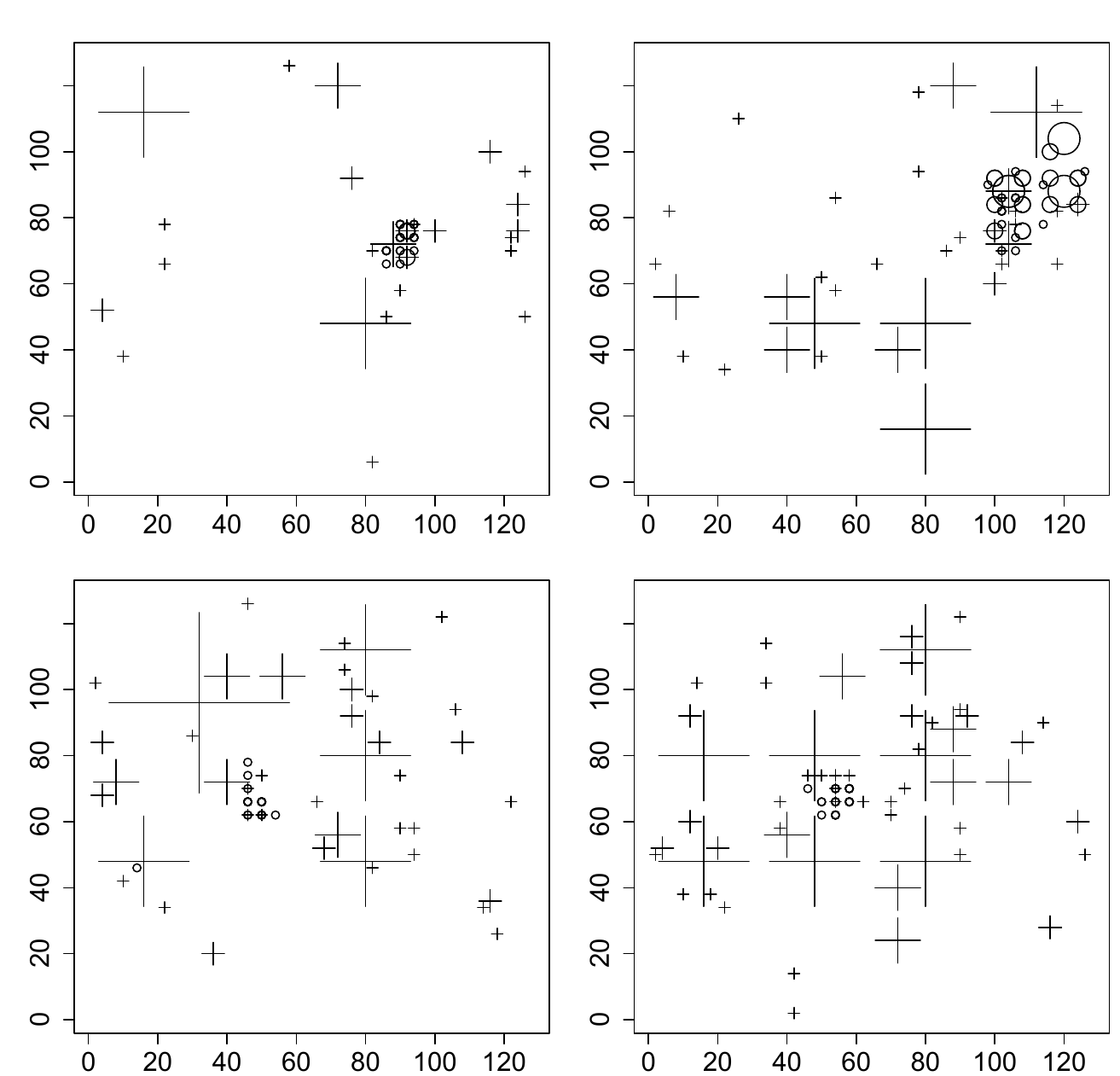}
\caption{Feature selection by $ESCV$ and $CV$ on four randomly selected voxels. The ``o"s represent features selected by both methods, while the ``+"s represent features selected only by $CV$. The axes represent the pixel location of the images. The position and size of the points represents the wavelet location and wavelet scale respectively. Note that most of the extra features $CV$ select are scattered and less biologically plausible.}
\label{Fig_fMRI_select}
\end{center}
\end{figure}

Now, each feature corresponds to a Gabor wavelet characterized by its location, frequency, and orientation. We plot the features selected by both $CV$ and $ESCV$ as well as the extra features selected by $CV$. 
The points in the plot represent the location and size of the Gabor wavelet selected. 
Figure \ref{Fig_fMRI_select} shows four randomly selected voxels. 

We can see quite clearly that the features selected by $ESCV$ are clustered in one area whereas the features selected by $CV$ but not $ESCV$ are scattered across the image. 
Biologically, we expect each voxel to respond only to a particular area of the visual receptive field.
This confirms that the extra features selected by $CV$ are most likely not meaningful.
Note that the location information of the Gabor wavelets were \emph{not} used in fitting the model.

\subsubsection{A Comparison with $CV$ for other choices of $V$}
\label{SecfMRI2}

The fMRI data set provides us with an unique opportunity. As seen in Table \ref{fMRITab} and Figure \ref{FigCors}, the predictive performance is similar despite the very different model size. Most of the Lasso solution path in this cas e, have comparable predictive performance. This is possible because we are using correlation as the prediction metric; scale is not scientifically important in this context. 

We compare the model sizes with $V=2$ and $V=5$. In this case, for each voxel, we repeat $V=2$ five times and $V=5$ twice and aggregate the results for the respective choice of $\lambda$. This is to bring the computation cost in line with the $V=10$ case. Table \ref{fMRITab2} gives the average model sizes by groups of 100 voxels. We see that lower $V$ does indeed correspond to a smaller model size. However, we note that even for $V=2$, the model size is still above that of $ESCV$ with the exception of the 100 voxels with the poorest predictive performance. We also note that there is relatively little change between $V=5$ and $V=10$, which bounds the common application of $V$-fold $CV$.

\begin{table}[htbp]
\begin{center}
\begin{tabular}{|c|c|c|c|c|}
\hline
Voxels  & \multicolumn{4}{c|}{Model Size}  \\
\cline{2-5}
 & $ESCV$ & $CV (V=2)$ & $CV (V=5)$ & $CV (V=10)$ \\
\hline
1-100 & 30.1 & 41.0 & 65.2 & 70.2 \\
101-200 & 27.0 & 36.0 & 55.7 &61.8 \\
201-300 & 22.6 & 26.9 & 43.4 & 49.6 \\
301-400 & 16.7 & 21.3 & 35.0 & 40.3 \\
401-500 & 16.5 & 14.4 & 27.3 & 33.6 \\
\hline 
\end{tabular}
\end{center}
\caption{Model size comparison on fMRI data set. The numbers are averaged across the respective hundred voxels. $CV$ continues to select larger models than $ESCV$ except in the 100 voxels with the poorest predictive performance.}
\label{fMRITab2}
\end{table}%

\subsection{Cytokine Data}
\label{SecCyt}
This data is from experiments performed by the Alliance for Cellular Signaling (AfCS), archived and made available at the Signaling Gateway, a comprehensive and free resource supported by the University of California, San Diego (UCSD). 
\citet{PraMauSub06} from the Bioinformatics and Data Coordination Laboratory at UCSD processed and analyzed this data in an attempt to identify signal pathways responsible for regulating cytokine release. There are 7 cytokines, 22 signal pathway predictors. The signal pathways cannot be directly manipulated. Instead, ligands are stimulated to elicit responses from the signal pathway predictors and cytokines. For each cytokine, we have about 100 samples, each corresponding to average measured responses of the cytokine and signal pathways when a specific ligand pair is stimulated.  

In the original study \citep{PraMauSub06}, principal component regression ($PCR$) is used to fit the data to a linear model and select the significant signal pathways. The selection is done by thresholding the estimated coefficients via a pseudo-bootstrap method. They do this for each of the seven cytokines. That is, they solve seven linear regression problems, each with $n \approx 100$ and $p = 22$, and apply thresholding to select the relevant signal pathways. These $PCR$ results are then merged with other data and analysis to derive a final minimal model ($MM$). 

\begin{figure}[htbp]
\begin{center}
\includegraphics[width=0.8\textwidth]{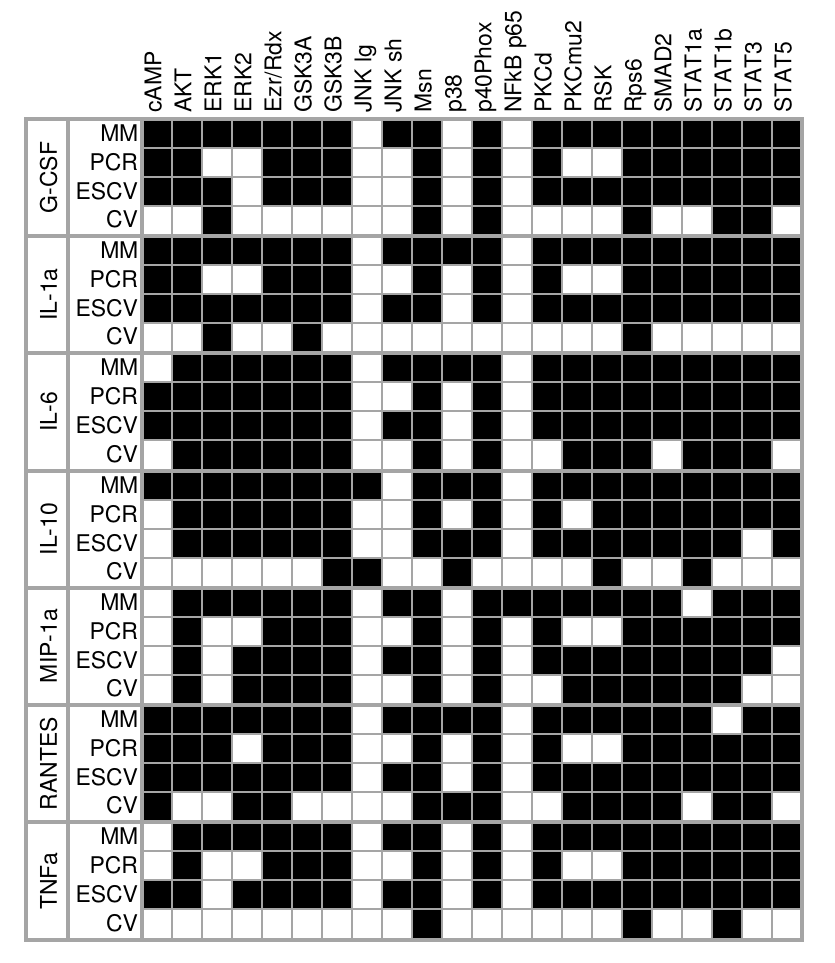}
\caption{Feature selection results on cytokine data. The columns represent signal pathways predictors and each block of four rows correspond to a cytokine. The four rows within each block represent the selections of the four methods:  the final minimal model ($MM$) and principal component regression ($PCR$) from the original study, and Lasso with $ESCV$ and $CV$. The white squares corresponds to selected predictors. With only one exception, $ESCV$ always selects the pathways that $MM$ (which we regard as ground truth) does, while having much smaller models than $CV$.}
\label{Fig_cytokine}
\end{center}
\end{figure}

We run Lasso with $ESCV$ and $CV$ on the seven linear regression problems and compare our results with the results from $PCR$ and $MM$. Fig \ref{Fig_cytokine} shows the feature selection results for the four methods. We regard $MM$
as the benchmark for feature selection performance
because it encompasses extra data and is not directly restricted by the linear model. 

We can see from Fig \ref{Fig_cytokine} that Lasso with $CV$ does poorly. It selects the most features for every cytokine, often by a large margin. Lasso with $ESCV$ on the other hand, selects the same or slightly larger number of features than $MM$. Moreover, with the exception of cytokine $TNFa$, $ESCV$ always includes the features $PCR$ selected which survived to the minimal model. 
In the case of $TNFa$, $PCR$ barely selects (close to threshold) the one feature that $ESCV$ missed.
$ESCV$ in general selects only about half the number of features $PCR$ selects.
There are far fewer false positives with respect to $MM$. At the same time, it rarely misses out any of the important features that $PCR$ picked up.

We stress again that $MM$ was derived using additional data independent of the seven linear regression problems we ran Lasso on. $ESCV$, in this case, has managed to extract more information from the limited linear regression data than $CV$ and $PCR$.

\section{Conclusion}
Regularization methods are employed to deal with problems in the increasingly common high dimensional setting.
However, the difficult problem of selecting the associated regularization parameter for interpretation or parameter estimation, is not well studied.
Our method $ESCV$ is based on estimation stability but also takes into account 
model fit via $CV$. 
With a similar parallelizable computational cost as $CV$, we have demonstrated that $ESCV$ is an effective alternative to the popular $CV$ for choosing the regularization parameter for the Lasso. 
On the whole, $ESCV$ is able to deliver comparable prediction performance as $CV$, and at the same time, do better in terms of other important statistical measures.
For the practical situation of dependent predictors, $ESCV$ has
an overall performance better than $CV$ for parameter estimation and
significantly outperforms $CV$ in model selection. 
In particular, we found much sparser models of less than half the size in both the real data sets 
from neuroscience and cell biology. These sparser $ESCV$ models preserves the prediction accuracy of the $CV$ models, and at the same time, are more parsimonious and are corroborated by  subject knowledge. 
We believe this result is not restricted to the Lasso but holds for other sparse regularization methods as well. 

We also believe that this method can also be readily extended to the classification problem through the generalized linear model, and leave this to future work.


\begin{table}[htbp]
\setlength{\tabcolsep}{3pt}
\begin{center}
\resizebox{.8\textwidth}{!}{%
\begin{tabular}{|c|c|c|c|c|c|c|c|c|c|c|c|c|c|}
\cline{3-14}
\multicolumn{2}{c}{} & \multicolumn{3}{|c|}{Estimation} & \multicolumn{3}{c|}{Prediction} & \multicolumn{3}{c|}{Model Selection} & \multicolumn{3}{c|}{Model Size}\\
\multicolumn{2}{c}{} & \multicolumn{3}{|c|}{error} & \multicolumn{3}{c|}{error} & \multicolumn{3}{c|}{$F$-measure} & \multicolumn{3}{c|}{} \\
\hline
$\rho$ & $\sigma$ & $ESCV$ & $CV$ & $BIC$ & $ESCV$ & $CV$ & $BIC$& $ESCV$ & $CV$& $BIC$ & $ESCV$ & $CV$& $BIC$\\
\hline
0 & 0.5 & 0.348 & \bf{0.301} & 0.342 & 0.348 & \bf{0.301} & 0.342 & 0.770 & 0.575 & \bf{0.771} & 16.0 & 24.8 & 16.0\\
0 & 1 & 0.656 & \bf{0.598} & 0.694 & 0.656 & \bf{0.598} & 0.694 & 0.699 & 0.565 & \bf{0.770} & 18.1 & 25.3 & 15.4\\
0 & 2 & 1.22 & \bf{1.16} & 1.84 & 1.22 & \bf{1.16} & 1.84 & \bf{0.595} & {0.560} & 0.402 & 18.2 & 22.5 & 3.41\\
\hline
0.2 & 0.5 & \bf{0.318} & 0.320 & 0.320 & 0.334 & \bf{0.324} & 0.336 & 0.716 & 0.635 & \bf{0.727} & 17.9 & 21.5 & 17.5\\
0.2 & 1 & \bf{0.576} & 0.590 & 0.587 & 0.559 & \bf{0.551} & 0.579 & 0.688 & 0.609 & \bf{0.711} & 19.0 & 22.8 & 18.0\\
0.2 & 2 & \bf{1.10} & 1.14 & 1.15 & \bf{1.07} & \bf{1.07} & 1.21 & 0.652 & 0.590 & \bf{0.685} & 18.2 & 21.6 & 15.2\\
\hline
0.5 & 0.5 & \bf{0.434} & 0.440 & 0.437 & \bf{0.421} & \bf{0.421} & 0.424 & 0.690 & 0.653 & \bf{0.700} & 18.9 & 20.6 & 18.5\\
0.5 & 1 & \bf{0.712} & 0.733 & 0.720 & \bf{0.553} & 0.557 & 0.568 & 0.668 & 0.621 & \bf{0.683} & 19.6 & 21.8 & 18.8\\
0.5 & 2 & \bf{1.30} & 1.36 & 1.32 & \bf{0.981} & 1.00 & 1.05 & 0.620 & 0.579 & \bf{0.636} & 18.4 & 20.7 & 16.8\\
\hline
0.9 & 0.5 & \bf{1.04} & \bf{1.04} & \bf{1.04} & 0.548 & \bf{0.546} & 0.552 & 0.637 & 0.636 & \bf{0.646} & 19.1 & 19.1 & 18.6\\
0.9 & 1 & \bf{1.48} & 1.51 & 1.49 & \bf{0.562} & 0.565 & 0.582 & 0.593 & 0.570 & \bf{0.598} & 18.2 & 19.5 & 17.3\\
0.9 & 2 & \bf{2.21} & 2.33 & 2.24 & \bf{0.767} & 0.78 & 0.844 & \bf{0.468} & 0.455 & 0.458 & 14.2 & 15.9 & 12.8\\
\hline
\end{tabular}}
\end{center}
\caption{Performance of $ESCV$, $CV$ and $BIC$ in picking the regularization parameter for the Lasso for $p=50$ with the base case Gaussian simulation $n=100$, constant correlation $\rho$.}
\label{Amb1Tab}

\begin{center}
\resizebox{.8\textwidth}{!}{%
\begin{tabular}{|c|c|c|c|c|c|c|c|c|c|c|c|c|c|}
\cline{3-14}
\multicolumn{2}{c}{} & \multicolumn{3}{|c|}{Estimation} & \multicolumn{3}{c|}{Prediction} & \multicolumn{3}{c|}{Model Selection} & \multicolumn{3}{c|}{Model Size}\\
\multicolumn{2}{c}{} & \multicolumn{3}{|c|}{error} & \multicolumn{3}{c|}{error} & \multicolumn{3}{c|}{$F$-measure} & \multicolumn{3}{c|}{} \\
\hline
$\rho$ & $\sigma$ & $ESCV$ & $CV$ & $BIC$ & $ESCV$ & $CV$ & $BIC$& $ESCV$ & $CV$& $BIC$ & $ESCV$ & $CV$& $BIC$\\
\hline
0 & 0.5 & 0.609 & \bf{0.541} & 0.803 & 0.609 & \bf{0.541} & 0.803 & 0.516 & 0.315 & \bf{0.711} & 28.5 & 53.5 & 16.1\\
0 & 1 & 1.15 & \bf{1.05} & 1.82 & 1.15 & \bf{1.05} & 1.82 & \bf{0.440} & 0.311 & 0.423 & 29.1 & 51.0 & 3.75\\
0 & 2 & 1.77 & \bf{1.74} & 2.16 & 1.77 & \bf{1.74} & 2.16 & \bf{0.304} & 0.292 & 0.0375 & 26.7 & 32.2 & 0.194\\
\hline
0.2 & 0.5 & \bf{0.553} & \bf{0.553} & 0.598 & 0.542 & \bf{0.534} & 0.620 & 0.418 & 0.367 & \bf{0.464} & 37.7 & 44.4 & 32.5\\
0.2 & 1 & \bf{0.995} & 1.01 & 1.45 & 0.932 & \bf{0.925} & 1.98 & 0.389 & 0.332 & \bf{0.426} & 39 & 47.8 & 16.5\\
0.2 & 2 & \bf{1.68} & 1.74 & 2.13 & \bf{1.58} & 1.60 & {3.42} & \bf{0.319} & 0.275 & 0.0879 & 34.5 & 43.1 & 0.88\\
\hline
0.5 & 0.5 & \bf{0.782} & 0.783 & 0.824 & 0.661 & \bf{0.658} & 0.752 & 0.392 & 0.376 & \bf{0.416} & 39.7 & 41.8 & 35.8\\
0.5 & 1 & \bf{1.22} & 1.23 & 1.33 & \bf{0.911} & 0.914 & 1.19 & 0.362 & 0.330 & \bf{0.387} & 40.1 & 45.2 & 32.1\\
0.5 & 2 & \bf{1.89} & 1.95 & 2.08 & \bf{1.40} & 1.42 & 3.35 & \bf{0.281} & 0.250 & 0.227 & 33.6 & 39.8 & 9.81\\
\hline
0.9 & 0.5 & \bf{1.64} & \bf{1.64} & 1.73 & 0.752 & \bf{0.750} & 0.848 & \bf{0.305} & \bf{0.305} & 0.303 & 37.2 & 37.3 & 32.8\\
0.9 & 1 & \bf{2.06} & \bf{2.06} & 2.14 & 0.767 & \bf{0.755} & 0.960 & \bf{0.243} & \bf{0.243} & 0.235 & 32.9 & 33.6 & 26.6\\
0.9 & 2 & \bf{2.61} & 2.71 & 2.64 & \bf{0.896} & 0.902 & 1.31 & \bf{0.139} & 0.133 & 0.123 & 22.7 & 26.6 & 16.2\\
\hline
\end{tabular}}
\end{center}
\caption{Performance of $ESCV$, $CV$ and extended $BIC$ in picking the regularization parameter for the Lasso for $p=500$ with the base case Gaussian simulation $n=100$, constant correlation $\rho$.}
\label{Amb2Tab}
\end{table}%

\begin{table}[h]
\setlength{\tabcolsep}{3pt}
\begin{center}
\resizebox{.8\textwidth}{!}{%
\begin{tabular}{|c|c|c|c|c|c|c|c|c|c|c|c|c|c|}
\cline{3-14}
\multicolumn{2}{c}{} & \multicolumn{3}{|c|}{Estimation} & \multicolumn{3}{c|}{Prediction} & \multicolumn{3}{c|}{Model Selection} & \multicolumn{3}{c|}{Model Size}\\
\multicolumn{2}{c}{} & \multicolumn{3}{|c|}{error} & \multicolumn{3}{c|}{error} & \multicolumn{3}{c|}{$F$-measure} & \multicolumn{3}{c|}{} \\
\hline
$\rho$ & $\sigma$ & $ESCV$ & $CV$ & $BIC$ & $ESCV$ & $CV$ & $BIC$& $ESCV$ & $CV$& $BIC$ & $ESCV$ & $CV$& $BIC$\\
\hline
0.3 & 0.5 & 0.498 & \bf{0.474} & 0.566 & 0.477 & \bf{0.439} & 0.557 & 0.537 & 0.377 & \bf{0.646} & 27.2 & 43.0 & 20.7\\
0.3 & 1 & 0.963 & \bf{0.933} & 1.46 & 0.920 & \bf{0.861} & 1.54 & 0.486 & 0.363 & \bf{0.569} & 28.7 & 43.8 & 9.73\\
0.3 & 2 & 1.65 & \bf{1.64} & 2.13 & 1.60 & \bf{1.54} & 2.34 & \bf{0.376} & 0.332 & 0.108 & 25.7 & 35.4 & 1.15\\
\hline
0.5 & 0.5 & 0.551 & \bf{0.544} & 0.609 & 0.463 & \bf{0.441} & 0.537 & 0.491 & 0.386 & \bf{0.576} & 30.6 & 41.7 & 24.3\\
0.5 & 1 & \bf{1.05} & \bf{1.05} & 1.51 & 0.880 & \bf{0.841} & 1.56 & 0.451 & 0.361 & \bf{0.503} & 31 & 42.8 & 11.3\\
0.5 & 2 & \bf{1.72} & 1.75 & 2.12 & 1.50 & \bf{1.45} & 2.45 & \bf{0.362} & 0.319 & 0.109 & 25.7 & 34.9 & 0.961\\
\hline
0.9 & 0.5 & \bf{1.11} & \bf{1.11} & 1.17 & 0.437 & \bf{0.429} & 0.496 & 0.420 & 0.392 & \bf{0.455} & 34.6 & 38.0 & 29.5\\
0.9 & 1 & \bf{1.77} & 1.83 & 1.88 & 0.713 & \bf{0.695} & 1.11 & 0.339 & 0.303 & \bf{0.351} & 30.6 & 36.6 & 18.2\\
0.9 & 2 & 2.33 & 2.54 & \bf{2.22} & 1.12 & \bf{1.08} & 2.18 & \bf{0.219} & 0.195 & 0.135 & 21.6 & 28.9 & 3.91\\
\hline
\end{tabular}}
\end{center}
\caption{Performance of $ESCV$, $CV$ and extended $BIC$ in picking the regularization parameter for the Lasso for the block correlation design. $n=100$, $p=300$.}
\label{BloTab}

\begin{center}
\resizebox{.8\textwidth}{!}{%
\begin{tabular}{|c|c|c|c|c|c|c|c|c|c|c|c|c|c|}
\cline{3-14}
\multicolumn{2}{c}{} & \multicolumn{3}{|c|}{Estimation} & \multicolumn{3}{c|}{Prediction} & \multicolumn{3}{c|}{Model Selection} & \multicolumn{3}{c|}{Model Size}\\
\multicolumn{2}{c}{} & \multicolumn{3}{|c|}{error} & \multicolumn{3}{c|}{error} & \multicolumn{3}{c|}{$F$-measure} & \multicolumn{3}{c|}{} \\
\hline
$\rho$ & $\sigma$ & $ESCV$ & $CV$ & $BIC$ & $ESCV$ & $CV$ & $BIC$& $ESCV$ & $CV$& $BIC$ & $ESCV$ & $CV$& $BIC$\\
\hline
0.5 & 0.5 & 0.537 & \bf{0.483} & 0.622 & 0.521 & \bf{0.461} & 0.610 & 0.557 & 0.363 & \bf{0.695} & 25.7 & 45.0 & 18.3\\
0.5 & 1 & 1.03 & \bf{0.946} & 1.55 & 1.01 & \bf{0.905} & 1.56 & 0.491 & 0.352 & \bf{0.548} & 26.6 & 45.1 & 8.29\\
0.5 & 2 & 1.68 & \bf{1.65} & 2.13 & 1.65 & \bf{1.61} & 2.19 & \bf{0.351} & 0.325 & 0.0767 & 26.3 & 33.7 & 1.47\\
\hline
0.9 & 0.5 & 0.788 & \bf{0.782} & 0.832 & 0.441 & \bf{0.425} & 0.492 & 0.479 & 0.402 & \bf{0.541} & 31.0 & 39.0 & 25.9\\
0.9 & 1 & \bf{1.38} & 1.39 & 1.61 & 0.816 & \bf{0.781} & 1.32 & 0.426 & 0.357 & \bf{0.472} & 29.7 & 38.8 & 14.8\\
0.9 & 2 & \bf{1.99} & 2.06 & 2.14 & 1.38 & \bf{1.33} & 2.41 & \bf{0.301} & 0.268 & 0.126 & 23.9 & 31.8 & 1.51\\
\hline
0.99 & 0.5 & \bf{1.91} & \bf{1.91} & 1.95 & 0.485 & \bf{0.482} & 0.535 & 0.324 & 0.322 & \bf{0.325} & 29.8 & 30.1 & 26.5\\
0.99 & 1 & \bf{2.31} & 2.32 & 2.34 & 0.568 & \bf{0.559} & 0.680 & \bf{0.233} & 0.226 & 0.229 & 25.1 & 26.5 & 20.7\\
0.99 & 2 & 2.70 & 2.82 & \bf{2.68} & 0.859 & \bf{0.858} & 1.15 & \bf{0.143} & 0.135 & 0.134 & 18.2 & 20.8 & 13.2\\
\hline
\end{tabular}}
\end{center}
\caption{Performance of $ESCV$, $CV$ and extended $BIC$ in picking the regularization parameter for the Lasso for the Toeplitz correlation design. $n=100$, $p=300$.}
\label{ToeTab}
\end{table}%

\begin{table}[h]
\setlength{\tabcolsep}{3pt}
\begin{center}
\small
\begin{tabular}{|c|c|c|c|c|c|c|c|c|c|c|c|c|c|}
\cline{3-14}
\multicolumn{2}{c}{}&\multicolumn{12}{|c|}{Constant correlation design}\\
\cline{3-14}
\multicolumn{2}{c}{} &\multicolumn{4}{|c}{$p=300$} &\multicolumn{4}{|c}{$p=50$}  &\multicolumn{4}{|c|}{$p=500$} \\
\cline{3-14}
\multicolumn{2}{c}{} & \multicolumn{2}{|c|}{True Positive} & \multicolumn{2}{c|}{False Positive} & \multicolumn{2}{c|}{True Positive} & \multicolumn{2}{c|}{False Positive} & \multicolumn{2}{c|}{True Positive} & \multicolumn{2}{c|}{False Positive}\\
\hline
$\rho$ & $\sigma$ & $ESCV$ & $CV$ & $ESCV$ & $CV$& $ESCV$ & $CV$ & $ESCV$ & $CV$& $ESCV$ & $CV$ & $ESCV$ & $CV$ \\
\hline
0 & 0.5 & 9.96 & 10.0 & 14.5 & 37.0& 10.0 & 10.0 & 5.96 & 14.8 & 9.91 & 9.99 & 18.5 & 43.5\\
0 & 1 & 9.00 & 9.68 & 17.3 & 37.2& 9.81 & 9.97 & 8.27 & 15.3 & 8.59 & 9.47 & 20.5 & 41.5\\
0 & 2 & 6.07 & 6.88 & 18.0 & 24.5& 8.38 & 9.09 & 9.78 & 13.4 & 5.57 & 6.16 & 21.1 & 26.0\\
\hline
0.2 & 0.5 & 9.98 & 9.98 & 21.7 & 27.7& 10.0 & 10.0 & 7.93 & 11.5 & 9.97 & 9.97 & 27.7 & 34.4\\
0.2 & 1 & 9.77 & 9.81 & 23.9 & 31.9& 9.98 & 9.99 & 9.01 & 12.8 & 9.53 & 9.6 & 29.5 & 38.2\\
0.2 & 2 & 7.65 & 7.92 & 22.6 & 30.3& 9.21 & 9.33 & 9.03 & 12.3 & 7.09 & 7.3 & 27.4 & 35.8\\
\hline
0.5 & 0.5 & 9.85 & 9.85 & 24.5 & 26.1& 9.97 & 9.97 & 8.92 & 10.6 & 9.75 & 9.75 & 30.0 & 32.1\\
0.5 & 1 & 9.41 & 9.43 & 26.1 & 30.9& 9.88 & 9.9 & 9.69 & 11.9 & 9.06 & 9.12 & 31.0 & 36.1\\
0.5 & 2 & 6.91 & 7.07 & 24.0 & 29.8& 8.81 & 8.9 & 9.59 & 11.8 & 6.11 & 6.23 & 27.5 & 33.5\\
\hline
0.9 & 0.5 & 7.79 & 7.80 & 25.2 & 25.2& 9.27 & 9.27 & 9.81 & 9.87 & 7.19 & 7.20 & 30.0 & 30.1\\
0.9 & 1 & 5.88 & 5.98 & 23.7 & 24.3& 8.35 & 8.43 & 9.83 & 11.1 & 5.21 & 5.30 & 27.7 & 28.3\\
0.9 & 2 & 2.87 & 3.06 & 17.9 & 21.1& 5.67 & 5.90 & 8.57 & 10.0 & 2.28 & 2.42 & 20.4 & 24.2\\
\hline
\end{tabular}

\vspace{1cm}
\begin{tabular}{|c|c|c|c|c|c|}
\cline{3-6}
\multicolumn{2}{c}{} & \multicolumn{4}{|c|}{Block design, $p=300$}\\
\cline{3-6}
\multicolumn{2}{c}{} & \multicolumn{2}{|c|}{True Positive} & \multicolumn{2}{c|}{False Positive}\\
\hline
$\rho$ & $\sigma$ & $ESCV$ & $CV$ & $ESCV$ & $CV$ \\
\hline
0.3 & 0.5 & 9.99 & 10.0 & 17.2 & 33.0\\
0.3 & 1 & 9.40 & 9.75 & 19.3 & 34.0\\
0.3 & 2 & 6.71 & 7.54 & 19.0 & 27.9\\
\hline
0.5 & 0.5 & 9.97 & 9.99 & 20.7 & 31.7\\
0.5 & 1 & 9.25 & 9.52 & 21.7 & 33.2\\
0.5 & 2 & 6.48 & 7.16 & 19.3 & 27.7\\
\hline
0.9 & 0.5 & 9.36 & 9.40 & 25.2 & 28.6\\
0.9 & 1 & 6.88 & 7.07 & 23.7 & 29.6\\
0.9 & 2 & 3.46 & 3.79 & 18.1 & 25.1\\
\hline
\end{tabular}
\qquad
\begin{tabular}{|c|c|c|c|c|c|}
\cline{3-6}
\multicolumn{2}{c}{} & \multicolumn{4}{|c|}{Toeplitz design, $p=300$}\\
\cline{3-6}
\multicolumn{2}{c}{} & \multicolumn{2}{|c|}{True Positive} & \multicolumn{2}{c|}{False Positive}\\
\hline
$\rho$ & $\sigma$ & $ESCV$ & $CV$ & $ESCV$ & $CV$ \\
\hline
0.5 & 0.5 & 9.94 & 10.0 & 15.8 & 35.0\\
0.5 & 1 & 8.98 & 9.70 & 17.6 & 35.4\\
0.5 & 2 & 6.38 & 7.10 & 19.9 & 26.6\\
\hline
0.9 & 0.5 & 9.83 & 9.85 & 21.2 & 29.1\\
0.9 & 1 & 8.44 & 8.70 & 21.2 & 30.1\\
0.9 & 2 & 5.10 & 5.59 & 18.8 & 26.2\\
\hline
0.99 & 0.5 & 6.45 & 6.46 & 23.4 & 23.6\\
0.99 & 1 & 4.08 & 4.12 & 21.0 & 22.4\\
0.99 & 2 & 2.02 & 2.08 & 16.1 & 18.7\\
\hline
\end{tabular}


\end{center}
\caption{Breakdown of the $F$-measure: the true positive and false positive rates of $ESCV$ and $CV$ for all the simulation scenarios. In all the cases above, there are 10 true variables and $p-10$ noise variables.}
\label{TPFPTab}
\end{table}%

\section*{Acknowledgements}

\thanks{
Lim would like to thank Jing Lei and Garvesh Raskutti for helpful discussions, and Yu
would like to thank Arnoldo Frigessi and Ingrid Glad for helpful comments.
This research is supported in part by US NSF grants DMS-1107000, SES-0835531
(CDI), NSF-CDS\&E-MSS grant 1228246, ARO grant W911NF-11-1-0114, and the Center for Science of
Information (CSoI), an US NSF Science and Technology Center, under grant
agreement CCF-0939370. 
}

\section*{Supplementary Materials}
\begin{description} 
\item[ESCV\_code.zip:] R code to perform the simulations described in the article. Refer to readme.txt for details.
\item[ESCV\_data.zip:] Data for the problems described in Section \ref{SecfMRI} and \ref{SecCyt}.
\end{description} 

\bibliography{ssrefs}

\bibliographystyle{asa}
\end{document}